\documentclass[twocolumn,nofootinbib]{revtex4}

\usepackage{dsfont}

\usepackage{graphicx}

\usepackage{enumitem}

\usepackage{xcolor}

\begin{document}

\title{On the value of the Immirzi parameter and the horizon entropy}

\author{C\'assio Pigozzo$^1$, Flora S. Bacelar$^2$, Saulo Carneiro$^{1,3}$}

\affiliation{$^1$Group of Gravitation and Cosmology, Instituto de F\'{\i}sica, Universidade Federal da Bahia, 40210-340, Salvador, BA, Brazil\\$^2$Group of Statistical Physics and Complex Systems, Instituto de F\'{\i}sica, Universidade Federal da Bahia, 40210-340, Salvador, BA, Brazil\\$^3$PPGCosmo, CCE, Universidade Federal do Esp\'irito Santo, 29075-910, Vit\'oria, ES, Brazil}

\date{\today}

\begin{abstract}
In Loop Quantum Gravity (LQG) the quantisation of General Relativity leads to precise predictions for the eigenvalues of geometrical observables like volume and area, up to the value of the only free parameter of the theory, the Barbero-Immirzi (BI) parameter. With the help of the eigenvalues equation for the area operator, LQG successfully derives the Bekenstein-Hawking entropy of large black holes with isolated horizons, fixing at this limit the BI parameter as $\gamma \approx 0.274$. In the present paper we show some evidence that a black hole with angular momentum $\hbar$ and Planck mass is an eigenstate of the area operator provided that $\gamma = \sqrt{3}/6 \approx 1.05 \times 0.274$. As the black hole is extremal, there is no Hawking radiation and the horizon is isolated. We also suggest that such a black hole can be formed in the head-on scattering of two parallel Standard Model neutrinos in the mass state $m_2$ (assuming $m_1 = 0$). Furthermore, we use the obtained BI parameter to numerically compute the entropy of isolated horizons with areas ranging up to $250\,l_P^2$, by counting the number of micro-states associated to a given area. The resulting entropy has a leading term ${\cal S} \approx 0.25\, {\cal A}$, in agreement to the Bekenstein-Hawking entropy. As the identification of the above eigenstate rests on the matching between classical areas and quantum area eigenvalues, we also present, on the basis of an effective quantum model for the Schwarzschild black hole recently proposed by Ashtekar, Olmedo and Singh, an expression for the quantum corrected area of isolated horizons, valid for any black hole mass. Quantum corrections are shown to be negligible for a Planck mass black hole, of order $10^{-3}$ relative to the classical area.
\end{abstract}

\maketitle

\section{Introduction}

As well known, the quantisation of gravity suffers, among others difficulties, from the non-convergence of its perturbative expansions, related to the absence of an adimensional coupling constant, contrary to what happens in other gauge theories \cite{woodard}. This has led to the development of non-perturbative approaches, among which Loop Quantum Gravity (LQG) is probably the most complete from a theoretical viewpoint \cite{lewandowski,thiemann}. Another difficulty is related to the absence of empirical facts that could drive the postulation of quantisation rules. In spite of that, LQG has successfully built consistent conjugate operators and their commutation relations, up to a free parameter that fixes a particular quantum representation, the Barbero-Immirzi (BI) parameter. Once this parameter is determined by any experiment, additional tests can rule out the theory, which in this way is falsifiable. The task is to find at least two independent tests in the realm of a so weak interaction and so small scales. Surprisingly enough, the theory has been confronted to the derivation of the horizon entropy of large black holes\footnote{For a discussion on the area quantisation of cosmological horizons, see e.g. \cite{mena,mena2} and references therein.}, explaining the linear relation between entropy and horizon area and fixing the BI parameter as $\gamma \approx 0.274$ in order to have the expected slope of $1/4$. This is done by counting the number of spin network configurations that generate a given horizon area. In this count the horizon is assumed isolated, as usually done for thermodynamic systems.

The main goal of this paper is to explore the possibility of an independent determination of the BI parameter. For that, we will initially identify a candidate for a physical eigenstate of the LQG area operator, constituted by a black hole of angular momentum $\hbar$, formed for instance by two interacting particles of spin $1/2$. The black hole must be extremal in order to avoid Hawking radiation, i.e. it has Planck mass $m_P$ and a Planck horizon radius $l_P$. Although it looks like a {\it gedanken} black hole, we will argue that it has an actual realisation if formed by two parallel neutrinos in the mass state $m_2$ (assuming the mass state $m_1 = 0$). Indeed, Dirac neutrinos carry magnetic moments due to vacuum fluctuations \cite{dipole}, and the dipoles repulsion energy at a distance $2l_P$ forms a horizon of radius $l_P$ and mass $m_P$ with $99.9\%$ precision \cite{saulo}. Both the found eigenstate and its physical realisation suggest that the classical expression for the horizon area is valid at the Planck scale, something that will be verified with the help of an effective model for spherically symmetric black holes. 

The BI parameter determined in this way differs $5\%$ from the approximate value derived from the entropy of large horizons. As a consistency test, we perform an exact counting of micro-states associated to small horizon areas running up to $250\, l_P^2$. The entropy ${\cal S}({\cal A})$ shows a linear leading term whose slope differs by less than $1\%$ from the Bekenstein-Hawking recipe. 

\section{A Planck scale black hole}

\label{Section II}

The simplest quantum black hole may be that formed in the scattering of two identical, repulsively interacting particles, at a centre-of-mass energy of the order of the Planck scale \cite{carr}. If the particles, for instance, have spin $1/2$ and carry parallel magnetic moments, the repulsion between the dipoles can lead to the formation of a Kerr black hole with angular momentum $\hbar$.\footnote{For a recent study of the classical Kerr solution in real Ashtekar variables, see \cite{olmedo}. Quantum black holes are discussed e.g. in \cite{olmedo2,perez,ashtekar,pranzetti}.} The formed horizon is isolated in the extremal case, when the surface gravity is zero and there is no Hawking radiation. In this case mass $M$, angular momentum $J$ and horizon radius $r_H$ are related by $a^2 = r_H^2 = J$, where $a = J/M$ \cite{BH}. This leads to $M = m_P$ and $r_H = l_P$. On the other hand, in the extremal limit the horizon area is reduced to
\begin{equation}\label{LQG1'}
{\cal A}  = 4\pi (r_H^2 + a^2) = 8\pi J.
\end{equation}
 For $J = 1$, it can be written as
\begin{equation} \label{LQG1}
{\cal A} = 8\pi \gamma l_P^2 \sum_{i=1}^4 \sqrt{j_i(j_i+1)},
\end{equation}
with $j_i = 1/2$ and $\gamma = \sqrt{3}/6$. This is the eigenvalues equation for the area operator of LQG \cite{smolin}, that fixes in this way the BI parameter. It can be interpreted as a horizon pierced by four spin network lines of colour $1/2$ or, equivalently, crossed by two lines, with two punctures per line \cite{saulo}.

Nevertheless, we still should find in Nature an actual physical system with the above features. It is noteworthy that it may indeed be formed in the head-on scattering of parallel neutrinos in a suitable mass state. Dirac neutrinos carry the smallest magnetic moment among the known particles of Standard Model, provided they have mass. Their magnetic dipole originates from vacuum fluctuations and its value involves the weak coupling constant, the fine structure constant and the masses of leptons and gauge bosons. At $1$-loop approximation and neglecting lepton masses as compared to gauge boson masses, the neutrino magnetic moment (in natural units) is given by \cite{dipole}
\begin{eqnarray}\label{dipoles4}
\mu_{\nu} &\approx& \frac{3eG_Fm_{\nu}}{8\sqrt{2}\pi^2},
\end{eqnarray}
where $G_F$ is the Fermi constant, $e$ is the elementary charge, and $m_{\nu}$ is the neutrino mass. Note that $2$-loops diagrams lead to relative corrections of order $\alpha/\pi \sim 10^{-3}$, where $\alpha$ is the low-energy fine structure constant. In fact, high order corrections to magnetic moments do not depend on the energy scale, involving only powers of $\alpha/\pi$ \cite{mandl}. Any scale dependence is absorbed, by renormalisation, in the term $e G_F m_{\nu}$ of Eq. (\ref{dipoles4}), that has dimension of charge/mass. This means that further vacuum polarisation effects are negligible compared to the classical interaction energy $U = \mu_{\nu}^2/r^3$, where $r$ is the neutrinos relative distance. In what follows, we shall assume that quantum gravity corrections are also negligible above the Planck length, an assumption that will be verified in Section IV. The next step is to take $r = 2 l_P$ and $U = M$, which leads to\footnote{$2l_P$ is the gravitational radius of an extremal Kerr black hole of event horizon radius $r_H = l_P$.}
\begin{eqnarray}\label{dipoles5}
M \approx \frac{9e^2G_F^2m_{\nu}^2}{1024\pi^4}.
\end{eqnarray}
If $M = m_P$, we have the Kerr solution described above, i.e. a physical eigenstate of the area operator. The no-hair conjecture, if valid at the Planck scale, assures that the magnetic dipoles are not observable from outside once the horizon is formed.

Although we do not know the absolute values of the neutrinos masses, flavor oscillation measurements give with precision the gaps between the squared masses. If we assume normal ordering and set the smallest mass $m_1 = 0$, the lightest massive state is $m_2 = (8.66 \pm 0.10) \times 10^{-3}$ eV ($1\sigma$) \cite{neutrinos,2020}. From (\ref{dipoles5}) we then have $M \approx 1.001\, m_P$. Reversing the argument, we would have an exact eigenstate if $m_2 \approx 8.654 \times 10^{-3}$ eV \cite{saulo}. The difference to the measured value has the same order of $2$-loops corrections to Eq. (\ref{dipoles4}). 

In spite of such a precision in the determination of both the BI parameter and neutrinos mass, matching classical horizon areas to the eigenvalues of the area operator is an {\it ad hoc} procedure that should be properly justified, as will be done in Section IV. On the other hand, our result for $\gamma$ must be independently confirmed through the evaluation of the horizon entropy, what we do in the next section.

\section{The horizon entropy}

The entropy of a black hole of horizon area ${\cal A}$ can be found by counting the number $N$ of spin network configurations that satisfy the eigenvalues equation of the area operator \cite{baez},
\begin{equation} \label{LQG}
{\cal A} = 8\pi \gamma l_P^2 \sum_{i=1}^n \sqrt{j_i(j_i+1)},
\end{equation}
where $n$ is the number of points on the horizon pierced by spin network lines of colours $j_i \in \mathds{Z}_+/2$. Furthermore, the condition of horizon isolation imposes to the punctures a set of second labels $m_i$ that must satisfy the ``projection constraint'' \cite{mitra,agullo,perez2}
\begin{equation}\label{projection}
\sum_{i=1}^{n} m_i = 0,
\end{equation}
with
\begin{equation}
m_i \in \{-j_i,-j_i+1,...,j_i-1,j_i\}.
\end{equation}
In the limit of large horizon areas, condition (\ref{LQG}) can be analytically solved up to terms that vanish in the limit ${\cal A} \rightarrow \infty$, leading to \cite{meissner}
\begin{equation}\label{Bekenstein}
{\cal S} = \ln N = \frac{\tilde{\gamma}}{4\gamma} {\cal A},
\end{equation}
where $\tilde{\gamma} \approx 0.274$ is the root of
\begin{equation}
1 = \sum_{k=1}^{\infty} (k+1) \exp \left( -\pi \tilde{\gamma} \sqrt{k(k+2)} \right).
\end{equation}
Eq. (\ref{Bekenstein}) fits the Bekenstein-Hawking entropy for $\gamma = \tilde{\gamma}$. It is also possible to show that (in the same limit of large areas) the projection constraint (\ref{projection}) does not affect the leading term (\ref{Bekenstein}).

For small black holes the entropy can be exactly evaluated with the help of appropriate generating functions \cite{agullo} or by a direct computation of all permitted micro-states for a given area \cite{corichi}. Although the former allows the generation of larger areas in shorter times, we have initially followed the latter procedure for simplicity, which has allowed us to obtain the number of states for areas running up to $160\, l_P^2$. The entropy was then found as ${\cal S} = \ln N$. The computation consisted of the following algorithm: 
\begin{enumerate}[label={(\roman*)}]

\item fix a value for ${\cal A}_0$ (in units of $l_P^2$); 

\item calculate the maximum number $n_{\text{max}}$ of punctures for which condition (\ref{LQG}) is satisfied, given by  the integer part of ${\cal A}_0/(4\pi \gamma \sqrt{3})$; 

\item generate all vectors [$j_i$] of length $n \leq n_{\text{max}}$ for which ${\cal A} \in ({\cal A}_0-\delta {\cal A}, {\cal A}_0+\delta {\cal A})$ for a chosen semi-interval $\delta {\cal A}$, excluding permutations of equals $j_i$; 

\item for each allowed vector, find all combinations [($j_i$,$m_i$)] satisfying the projection constraint (\ref{projection}); 

\item vary ${\cal A}_0$ from ${\cal A}_{\text{min}}$ to ${\cal A}_{\text{max}}$ with a chosen step.

\end{enumerate}
The prohibition of permutations with equals $j_i$ comes from the indistinguishability of punctures with equal labels ($j_i,m_i$). Therefore, a given vector [$j_i$] of length $n$ with $n_s$ elements $j_s$ will have multiplicity
\begin{equation}
\frac{n!}{\prod_s (n_s!)}.
\end{equation}
If we do not impose the projection constraint, each vector will also have an additional multiplicity
\begin{equation}
\prod_{s} (2j_s+1)^{n_s}.
\end{equation}

\begin{figure}
 \begin{center}
 \includegraphics[width=0.5\textwidth]{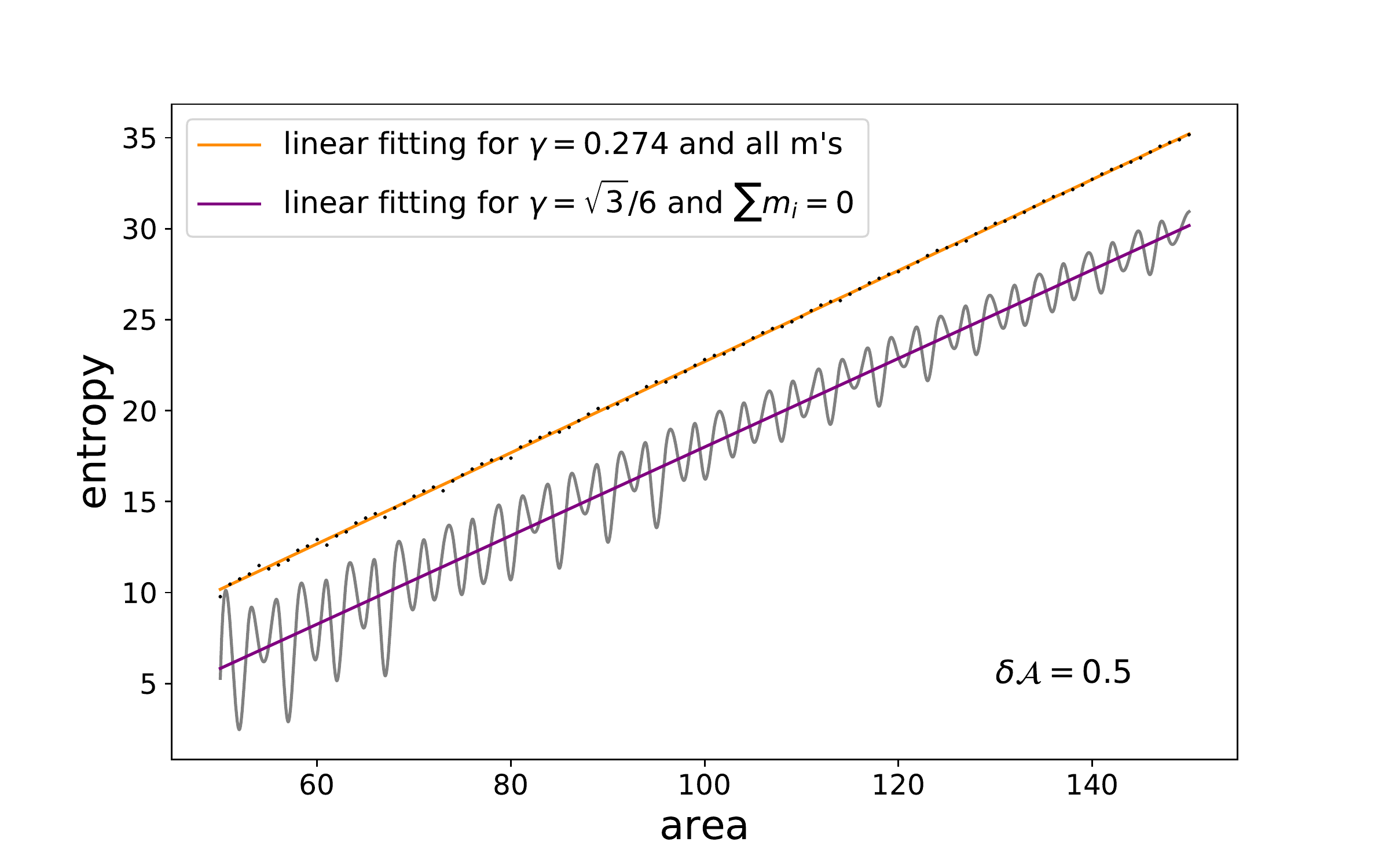}
 \end{center}
 \vspace{-0.3in}
 \caption{${\cal S}$ $\times$ ${\cal A}$ for $\delta {\cal A} = 0.5\, l_P^2$, without the projection constraint for $\gamma = 0.274$ (orange) and with the projection constraint for $\gamma = \sqrt{3}/6$ (purple).}
 \label{plot1}
 \end{figure}

We fixed $\delta {\cal A} = 0.5\, l_P^2$ as in \cite{corichi} and, to avoid superposition of intervals, varied ${\cal A}_0$ in steps of $2 \delta{\cal A}$. Without imposing the projection constraint, and taking $\gamma = 0.274$, we reproduced the findings of \cite{corichi}, i.e. a linear relation between ${\cal S}$ and ${\cal A}$ with slope $0.2504$, in excellent agreement with the large area approximation. The result is the orange curve shown in Fig.\,\ref{plot1}. When we impose the projection constraint and still take $\gamma = 0.274$, the oscillatory behaviour found in \cite{agullo,corichi} is evidenced. The fitting of a straight line gives a slope $0.254$, in contrast to the result $0.237$ reported in \cite{corichi}. Actually, the computation with the projection constraint is sensitive to the adopted semi-interval $\delta {\cal A}$ and to the step of variation of ${\cal A}_0$. With $\gamma = \sqrt{3}/6$ and without including the projection constraint, a linear relation is recovered with slope $0.238$, in agreement to the analytic approximation (\ref{Bekenstein}). When we include the projection constraint the oscillations reappear and the fitting of a straight line gives a slope $0.243$. The corresponding curve (in purple) is also shown in Fig.\,\ref{plot1}.  One can see that the oscillations are attenuated for larger areas, as expected in the thermodynamic limit. Note as well that, with or without the projection constraint, the slopes for $\gamma = 0.274$ are $1.05$ times higher than for $\gamma = \sqrt{3}/6$, in accordance to the large areas expression (\ref{Bekenstein}) and to the ratio between these values of the BI parameter.

\begin{figure*}
 \begin{center}
 \includegraphics[width=0.96\textwidth]{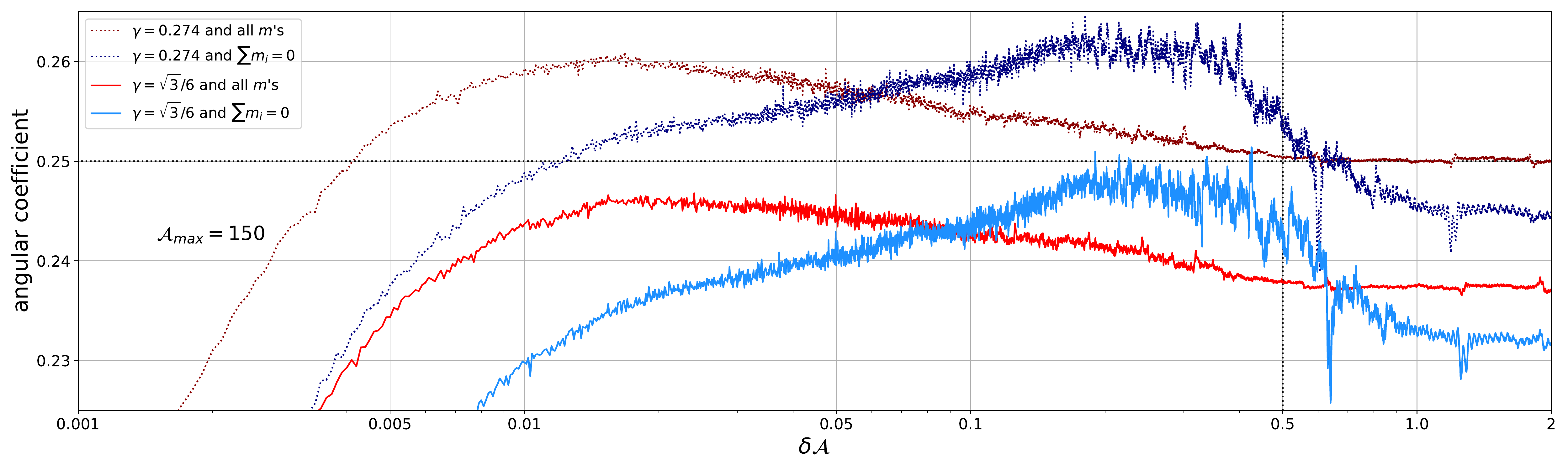}
 \end{center}
  \begin{center}
 \includegraphics[width=0.96\textwidth]{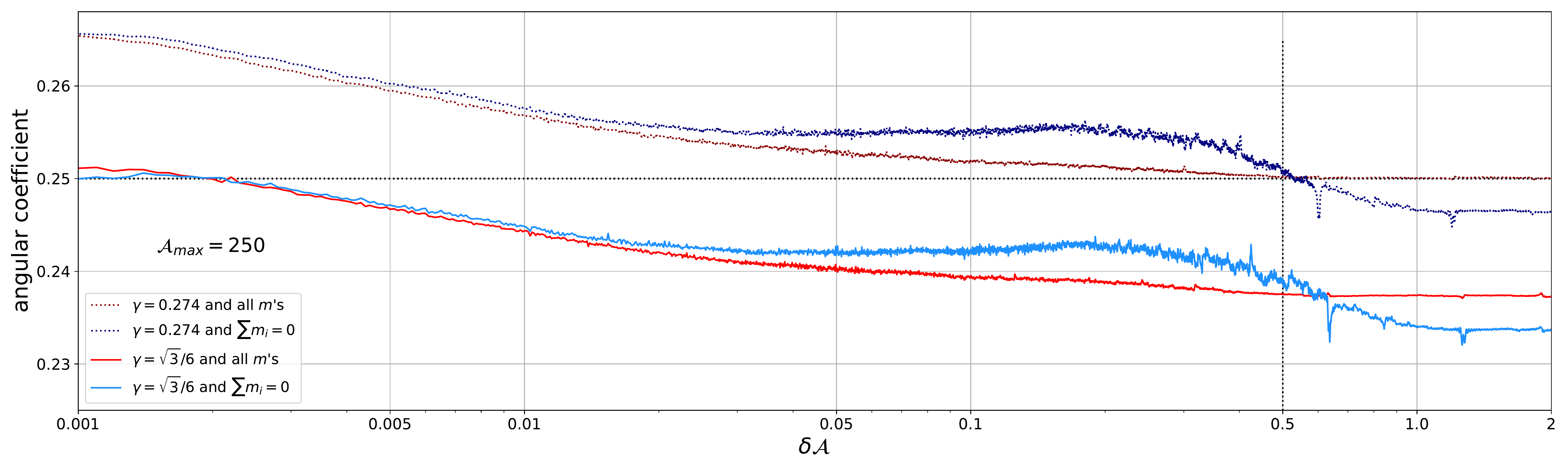}
 \end{center}
 \vspace{-0.3in}
 \caption{{\bf Upper panel:} ${\cal S}$ $\times$ ${\cal A}$ slope as a function of $\delta {\cal A}$ for $\gamma = \sqrt{3}/6$ and $\gamma = 0.274$, with and without the projection constraint, for ${\cal A}_{\text{max}} = 150\, l_P^2$. {\bf Lower panel:} The same for ${\cal A}_{\text{max}} = 250\, l_P^2$.}
 \label{plot2}
 \end{figure*}

We refined the analysis for $\gamma = \sqrt{3}/6$ varying the semi-interval $\delta {\cal A}$ in steps of $10^{-4}\, l_P^2$, for ${\cal A}_0$ running from $50\, l_P^2$ to $150\, l_P^2$, and we have found the correlations shown in the upper panel of Fig.\,\ref{plot2} between the ${\cal S} \times {\cal A}$  slope and the adopted area bin. Without the projection constraint (red line), the slope grows for smaller bins, approaching a maximum around $\delta {\cal A} = 0.02\, l_P^2$, for which  the angular coefficient is $0.246$. When the projection constraint is imposed (blue line), the slope approaches a maximum around $\delta {\cal A} = 0.25\, l_P^2$. At this point the angular coefficient is $0.249$, less than $1\%$ below the Bekenstein-Hawking value. The correspondent results for $\gamma = 0.274$ are also shown in Fig.\,\ref{plot2}. As expected, the curves are shifted by a factor of $\approx 1.05$, with a maximum slope of $\approx 0.263$. The correlations found may possibly be better understood with the help of the counting presented in \cite{agullo}, based on number theory and combinatory methods. Nevertheless, let us comment that for large area bins the fine grain structure of micro-states distribution is lost even for the smallest areas, which could explain the low values obtained for the slope with $\delta {\cal A} \gtrsim 0.5\, l_P^2$, that coincide with the large areas approximation (\ref{Bekenstein}) when the projection constraint is not considered. On the other hand, for small area bins the entropy oscillations become pronounced even for the largest areas, allowing in this way the fitting of straight lines with lower inclinations. The competition between these two effects, that are stronger when the projection constraint is imposed, may explain the maximum around $\delta {\cal A} \approx 0.25\, l_P^2$ and the oscillatory pattern observed.

If this interpretation is correct, we would expect a drift of the maximum slope to smaller area bins when larger maximum areas are used. In order to verify it, we have extended our computations up to ${\cal A}_{\text{max}} = 250\, l_P^2$. In this step the computation time was significantly shortened by using a formula for the $m$-degeneracy given by \cite{agullo}
\begin{equation}
\frac{1}{L} \left[ \prod_{i=1}^{n} (2j_i + 1) + \sum_{l=1}^{L-1} \prod_{i=1}^{n} \frac{\sin[2\pi(2j_i+1)l/L]}{\sin(2\pi l/L)} \right],
\end{equation}
where $L = 1 + 2 \sum j_i$ if $\sum j_i$ is integer, and $L = 2 +  2\sum j_i$ if $\sum j_i$ is half-integer.
Results are shown in the lower panel of Fig.\,\ref{plot2}.  Now the maximum slope is approached at $\delta {\cal A} \approx 0.002\, l_P^2$, given again by $\approx 0.25$ for $\gamma = \sqrt{3}/6$ when the projection constraint is imposed. For $\gamma = 0.274$ the maximum slope is again $\approx 0.263$, which suggests that the large area estimation of $\gamma$ presents a $5\%$ imprecision.

\begin{figure}
\begin{center}
\includegraphics[width=0.45\textwidth]{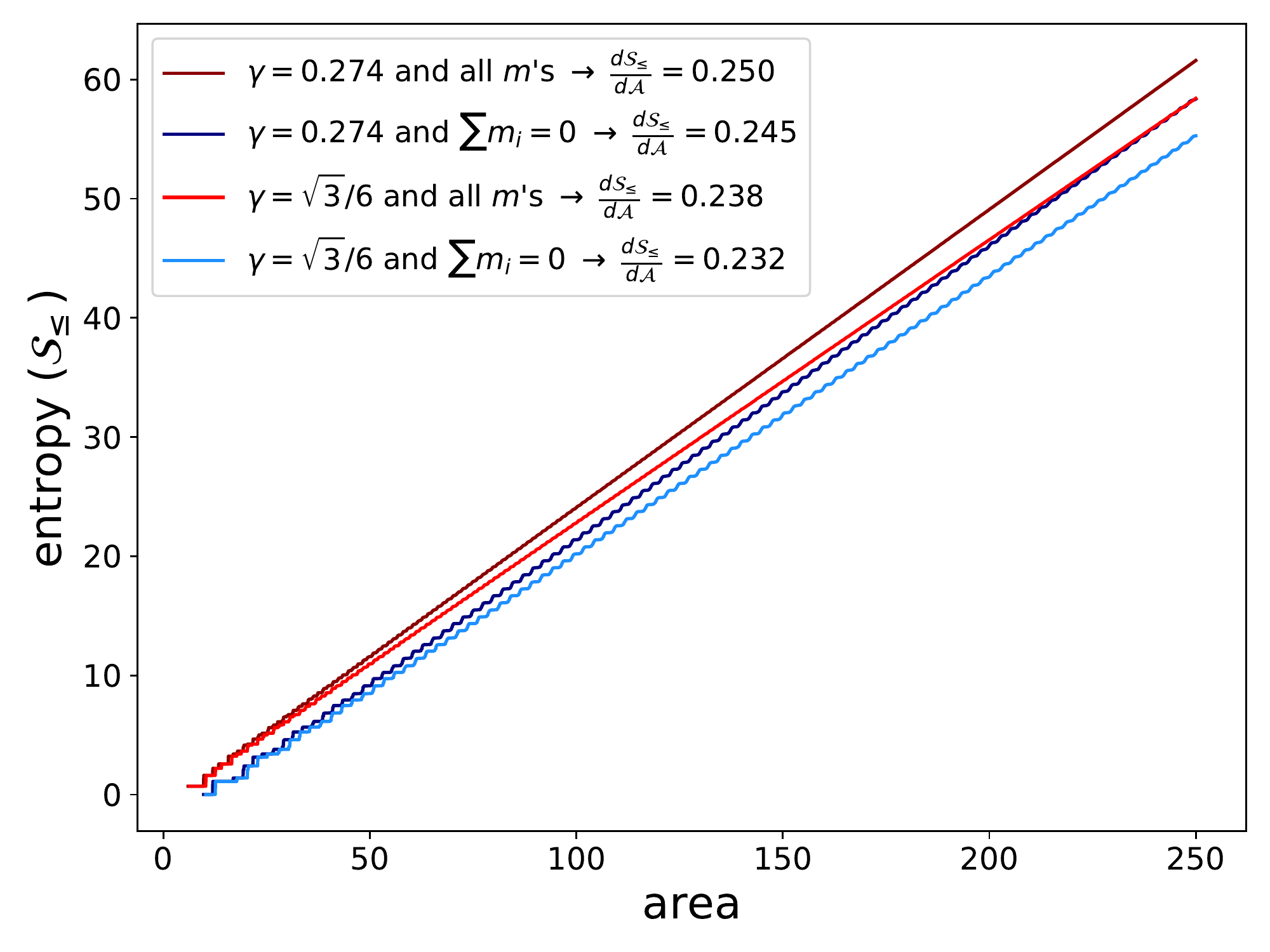}
\end{center}
 \vspace{-0.3in}
 \caption{${\cal S}_{\leq}$ $\times$ ${\cal A}$, with and without the projection constraint, for $\gamma = 0.274$ and $\gamma = \sqrt{3}/6$. The slopes found reproduce those shown in Fig.\,\ref{plot2} for the largest values of $\delta {\cal A}$.}
\label{plot3}
 \end{figure}

For completeness, we have also computed the entropy as defined in \cite{agullo}, namely
\begin{equation}
{\cal S}_{\leq} ({\cal A}) = \ln\, [1 + N_{\leq} ({\cal A})],
\end{equation}
where $N_{\leq} ({\cal A})$ is the number of vectors $[j_i]$ that generate areas in the interval $(0, {\cal A}]$. This definition is appropriate in the thermodynamic limit of large areas, and we have reproduced the slopes shown in Fig.\,\ref{plot2} for the largest values of $\delta {\cal A}$. The resulting ${\cal S}_{\leq} \times {\cal A}$ plots present the characteristic stair profile found in \cite{agullo}, as shown in Fig.\,\ref{plot3}.

\section{The horizon area}

Our goal in this section is to understand how can the classical horizon (\ref{LQG1'}) correspond to the area operator eigenvalue (\ref{LQG1}), contrary to our intuitive belief that quantum corrections to horizon areas should not be negligible at Planck scales. Using an effective, LQG inspired solution for a Schwarzschild quantum black hole \cite{ashtekar,Olmedo2}, we shall obtain an analytic expression for the quantum-corrected horizon area, and we will show that, for a black hole with Planck mass, the relative correction to the classical area is $\sim 10^{-3}$. Of course, in this case the Hawking radiation backreaction cannot be neglected, as the Hawking temperature is $\approx 4\%$ of the black hole mass. Nevertheless, it has no role at all if the horizon is assumed isolated, and we would not be extrapolating too much in assuming that quantum corrections are also negligible for a Planck mass Kerr black hole.

\subsection{Non-isolated horizons}

The interior of the spherically symmetric black hole of Ashtekar, Olmedo and Singh (AOS) is isometric to a vacuum Kantowski-Sachs spacetime with metric given, in Kruskal coordinates, by\footnote{See \cite{mariam} for a critical discussion on the asymptotic structure of the AOS solution.}
\begin{equation} \label{KS}
ds^2 = - N_{\tau}^2 d\tau^2 + \frac{p_b^2}{p_c L_o^2} dx^2 + p_c (d\theta^2 + \sin^2\theta d\phi^2).
\end{equation}
Here, $p_b$ and $p_c$ are canonical momenta conjugate to the respective configuration variables $b$ and $c$ of the dynamical space, and $L_o$ is a (non-observable) infrared cut-off introduced to avoid divergences in the Hamiltonian constraint. The lapse $N_{\tau}$ and the black hole mass are
\begin{equation} \label{lapse}
N_{\tau} = \frac{\gamma \sqrt{p_c} \delta_b}{\sin(\delta_b b)},
\end{equation}
\begin{equation}  \label{m1}
m = \frac{\sin(\delta_c c) p_c}{\gamma L_o \delta_c},
\end{equation}
where the constants of motion $\delta_b$ and $\delta_c$ encode quantum geometry corrections to the classical metric. The solution of the effective dynamical equations is given by
\begin{equation} \label{2.22}
p_c = 4m^2 \left( e^{2T} + \frac{\gamma^2 L_0^2 \delta_c^2}{64m^2} e^{-2T} \right),
\end{equation}
\begin{equation} \label{2.25}
p_b = - \frac{2\sin(\delta_c c)}{\delta_c} \frac{\sin(\delta_b b)}{\delta_b} \frac{p_c}{\gamma^2 + \frac{\sin^2(\delta_b b)}{\delta_b^2}},
\end{equation} \label{b}
\begin{equation} \label{2.23}
\cos(\delta_b b) = b_o \tanh \left[ \frac{b_o T}{2} + \tanh^{-1} \left( \frac{1}{b_o} \right) \right],
\end{equation}
with
\begin{equation}
b_o \equiv \sqrt{1 + \gamma^2 \delta_b^2},
\end{equation}
plus an additional equation for $c$ consistent with (\ref{m1}). The time variable $T$ is defined by $\tau = 2m e^{T}$, with $T = 0$ corresponding to the event horizon. The conjugate momentum $p_c$ presents a minimum at a transition surface ${\cal T}$ from which the trapped black hole interior tunnels to an anti-trapped white hole solution, resolving in this way the classical singularity. This minimum is given by
\begin{equation} \label{Pc}
p_c|_{\cal{T}} = m \gamma L_0 \delta_c,
\end{equation}
and it occurs at
\begin{equation} \label{time}
T_{\cal T} = \frac{1}{2} \ln \left( \frac{\gamma L_o \delta_c}{8m} \right).
\end{equation}

The strategy of the AOS effective model consists of matching the minimal areas enclosed by holonomies on the transition surface ${\cal T}$ to the LQG area gap $4\pi \sqrt{3} \gamma l_P^2$. Assuming that the links of these minimal plaquettes are proportional to $\delta_b$ on the $\theta$-$\phi$ $2$-spheres and to $\delta_c$ along the $x$ direction, we obtain \cite{ashtekar}
\begin{equation} \label{4.5}
\alpha^2 \delta_b^2 p_c|_{\cal{T}} = \sqrt{3} \gamma,
\end{equation}
\begin{equation} \label{4.4}
\alpha^2 \delta_c \delta_b |p_b||_{\cal{T}} = 2\sqrt{3} \gamma,
\end{equation}
where $\alpha > 0$ is a factor of proportionality with the order of unity\footnote{In Ref. \cite{ashtekar} $\alpha$ is assumed 1. As we will see, this assumption is too restrictive for Planck size horizons.}. 
From (\ref{Pc}) and (\ref{4.5}) we have
\begin{equation} \label{1}
L_0 \delta_c = \frac{\sqrt{3}}{m\alpha^2 \delta_b^2}.
\end{equation}
Substituting this result into (\ref{4.4}), it follows that\footnote{We choose $p_b < 0$ in accordance to the classical solution of the dynamical equations (see footnote 6 of \cite{ashtekar}).} 
\begin{equation} \label{2}
p_b|_{\cal{T}} = -2 m \gamma L_0 \delta_b.
\end{equation}
Using (\ref{2}) and (\ref{Pc}) in (\ref{2.25}) we obtain
\begin{equation} \label{3}
\delta_b^2 =  \frac{\sin (\delta_c c) \sin(\delta_b b)}{\gamma^2 + \frac{\sin^2(\delta_b b)}{\delta_b^2}} \quad \quad (\text{on}\; \cal{T}).
\end{equation}
On the other hand, from (\ref{Pc}) and (\ref{m1}) we see that
\begin{equation} \label{4}
\sin (\delta_c c)|_{\cal T} = 1,
\end{equation}
and, therefore,
\begin{equation} \label{5}
\delta_b = \frac{\frac{\sin(\delta_b b)}{\delta_b}}{\gamma^2 + \frac{\sin^2(\delta_b b)}{\delta_b^2}} \quad \quad (\text{on}\; \cal{T}).
\end{equation}
Let us now define
\begin{equation} \label{6}
x \equiv \frac{\sin (\delta_b b)}{\delta_b}.
\end{equation}
Eq. (\ref{5}) is rewritten as
\begin{equation} \label{7}
\delta_b x^2 - x + \delta_b \gamma^2 = 0,
\end{equation}
with roots
\begin{equation} \label{8}
x = \frac{1 \pm \sqrt{1 - 4 \delta_b^2 \gamma^2}}{2\delta_b},
\end{equation}
which leads to
\begin{equation} \label{8b}
\sin (\delta_b b) = \frac{1 \pm \sqrt{1 - 4 \delta_b^2 \gamma^2}}{2} \quad \quad (\text{on}\; \cal{T}).
\end{equation}

\begin{figure}
 \begin{center}
\includegraphics[width=0.4\textwidth]{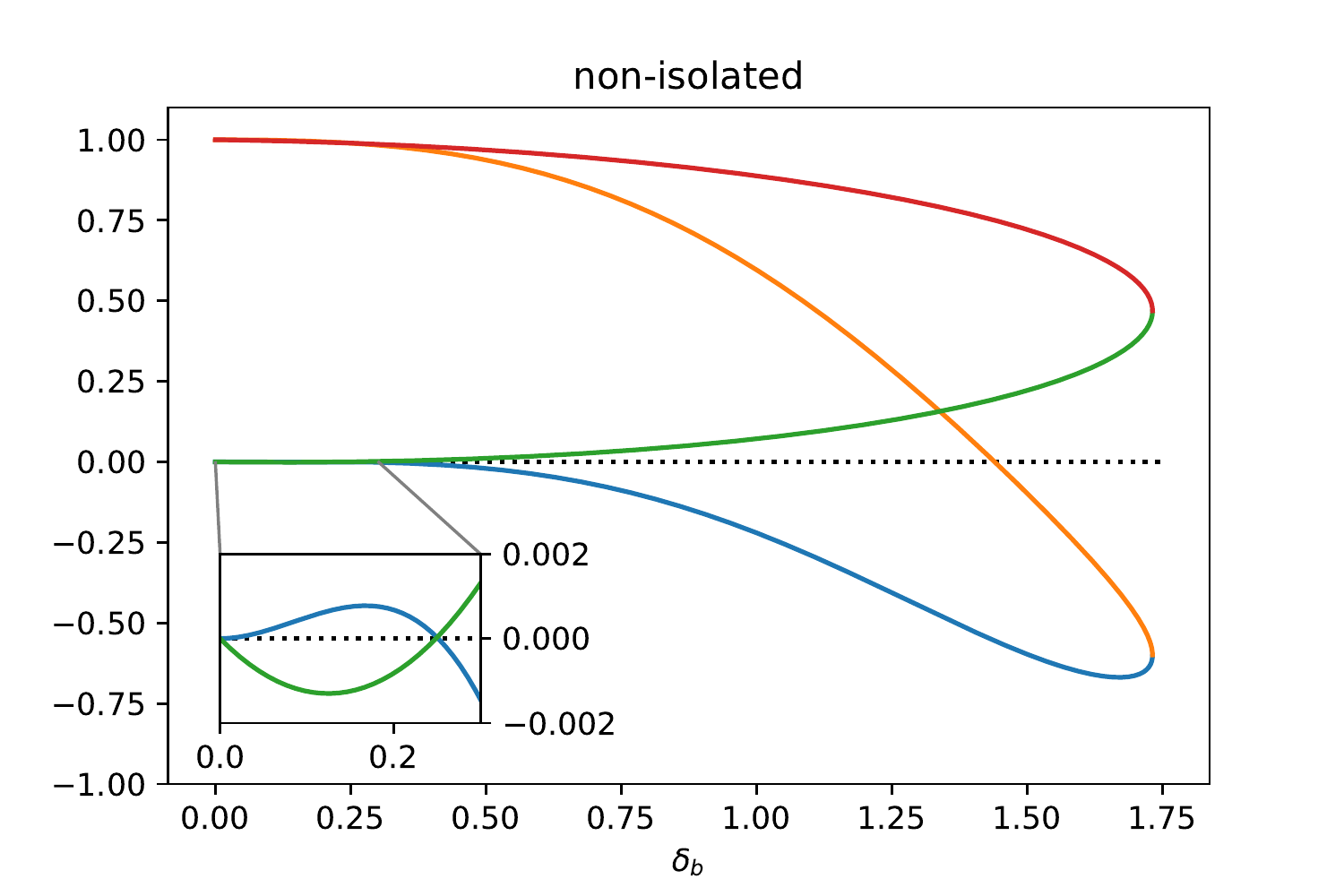}
\includegraphics[width=0.4\textwidth]{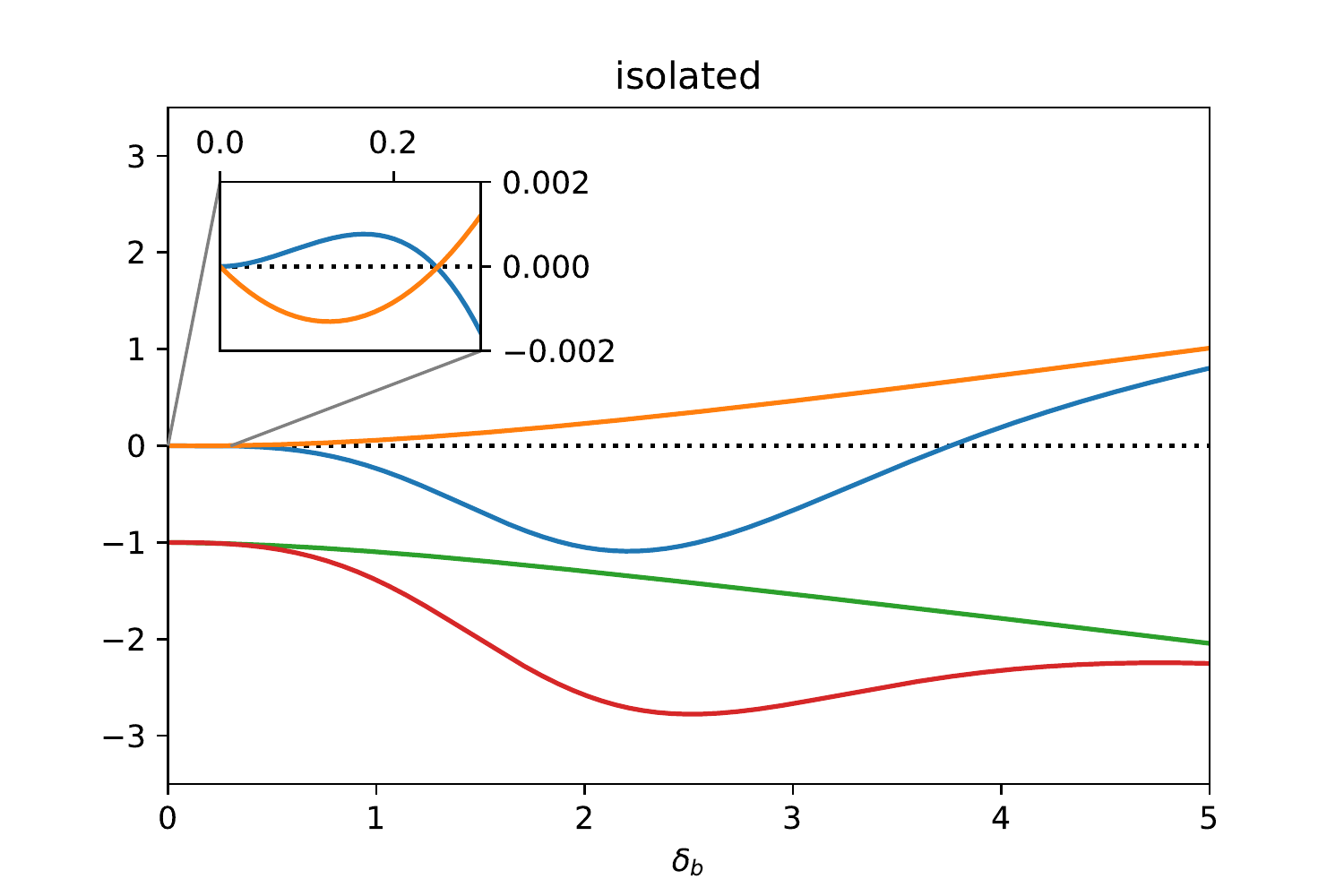}
 \end{center}
 \caption{Numerical solutions for $\delta_b$ in the non-isolated horizon (upper panel) and isolated horizon (lower panel) cases, for $m\alpha = 1$. The solutions (in the horizontal axes) correspond to the intersections of the lines $y = 0$.}
 \label{plot1'}
 \end{figure}

Now, if we take Eq.~(\ref{2.22}) at $T = 0$ and use (\ref{1}), we find the horizon area
\begin{equation} \label{horizon}
{\cal A} = 4\pi p_c = 16 \pi m^2 \left( 1 + \frac{3 \gamma^2}{64 m^4 \alpha^4 \delta_b^4} \right).
\end{equation}
In the large mass limit we recover the classical Schwarzschild result.
For small $m$, the correction to the horizon area depends on the value of $\delta_b$ along the correspondent dynamical trajectory, which is determined by simultaneously solving Eqs. (\ref{2.23}) and (\ref{8b}), using, in the former, Eqs. (\ref{time}) for $T$ and (\ref{1}) for $L_o \delta_c$.  If we take $m\alpha = 1$ and $\gamma = \sqrt{3}/6$, the largest root found in a numerical integration is $\delta_b \approx 1.44$, leading to a relative correction to the horizon area of $\approx 9 \times 10^{-4}$. The result of the numerical integration is shown in the upper panel of Fig.~\ref{plot1'}. The four branches refer to the possible combinations of the two branches of $\tanh^{-1}$ in (\ref{2.23}) and the two signs in (\ref{8b}). The solutions for $\delta_b$ (horizontal axis) correspond to the intersections of the line $y = 0$.

\begin{figure}
 \begin{center}
\includegraphics[width=0.4\textwidth]{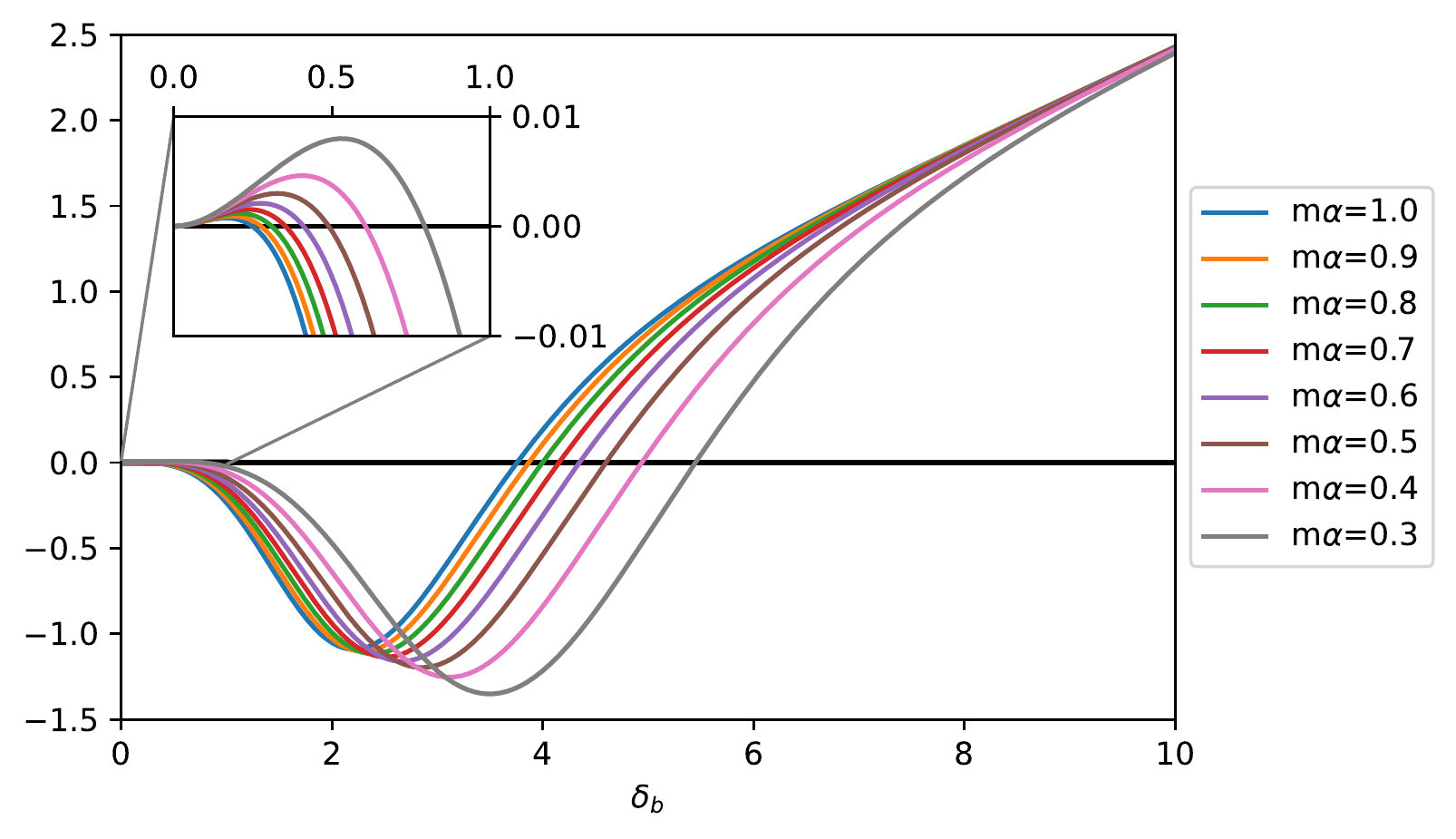}
\includegraphics[width=0.4\textwidth]{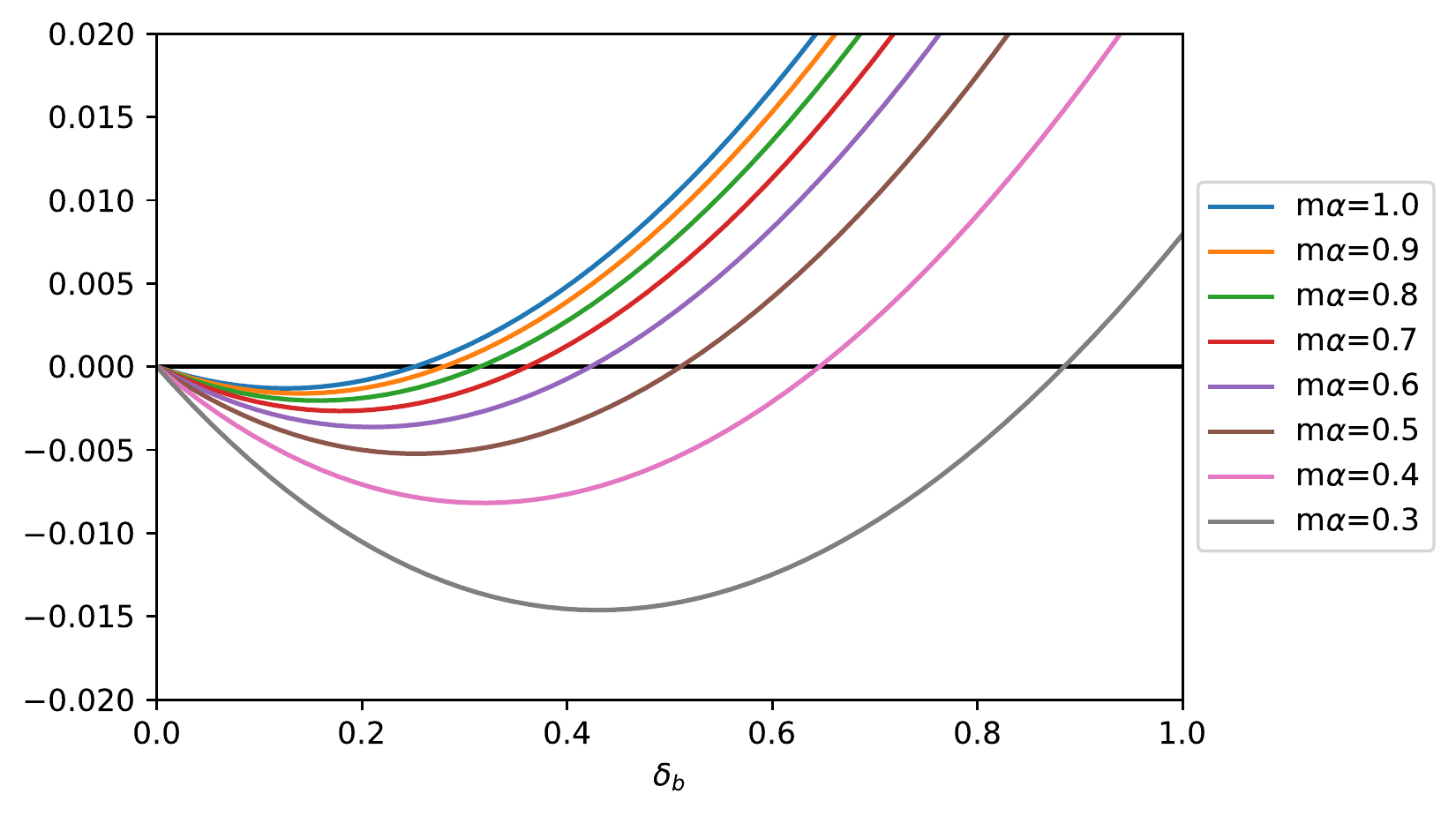}
 \end{center}
 \caption{Numerical solutions for $\delta_b$ in the isolated horizon case for black hole masses in the interval $m \alpha \in [0.3,1]$. The solutions (in the horizontal axes) correspond to the intersections of the lines $y = 0$. The panels correspond to the two branches with real solutions.}
 \label{plot2'}
 \end{figure}
 
  \begin{figure}
 \begin{center}
\includegraphics[width=0.4\textwidth]{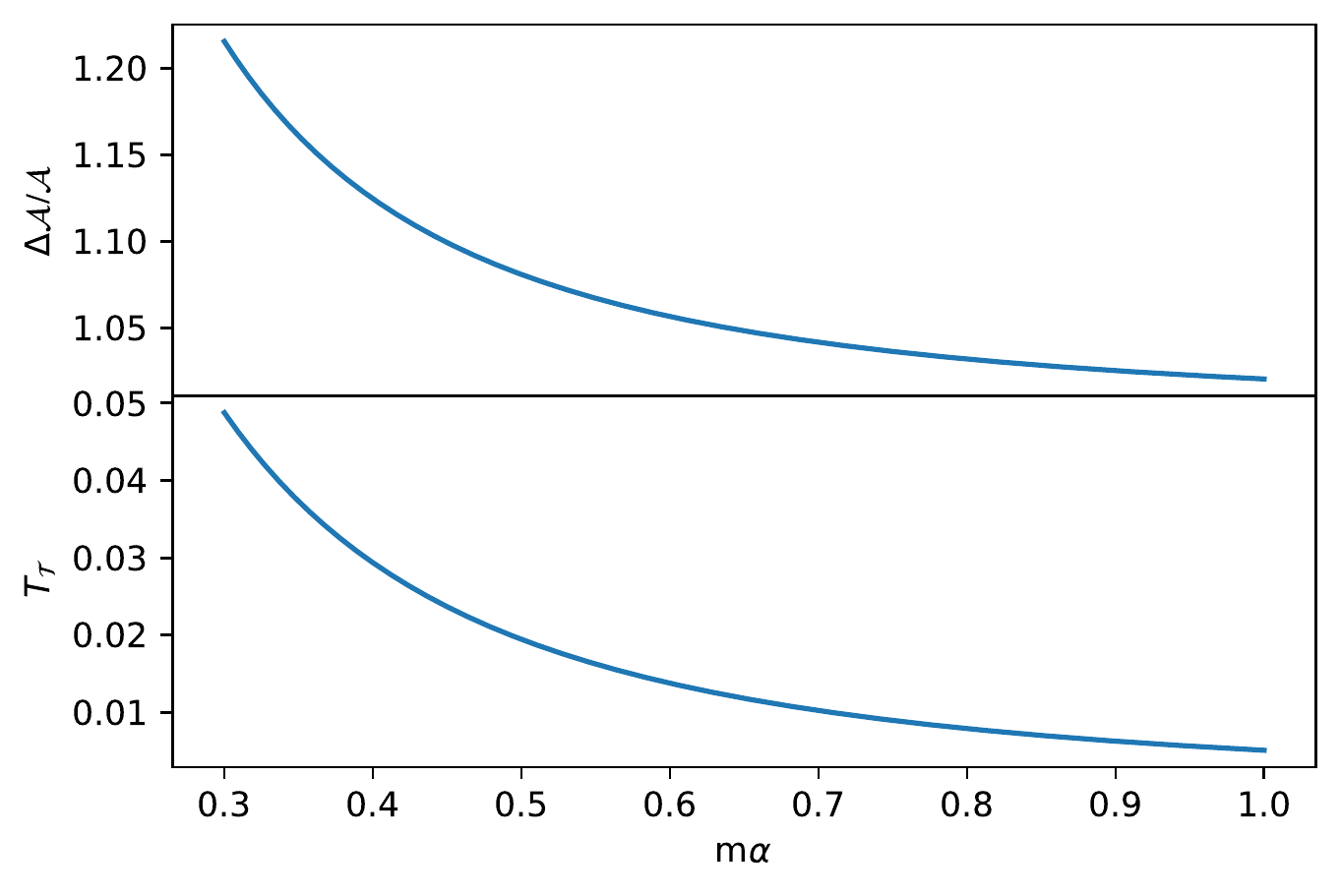}
\includegraphics[width=0.4\textwidth]{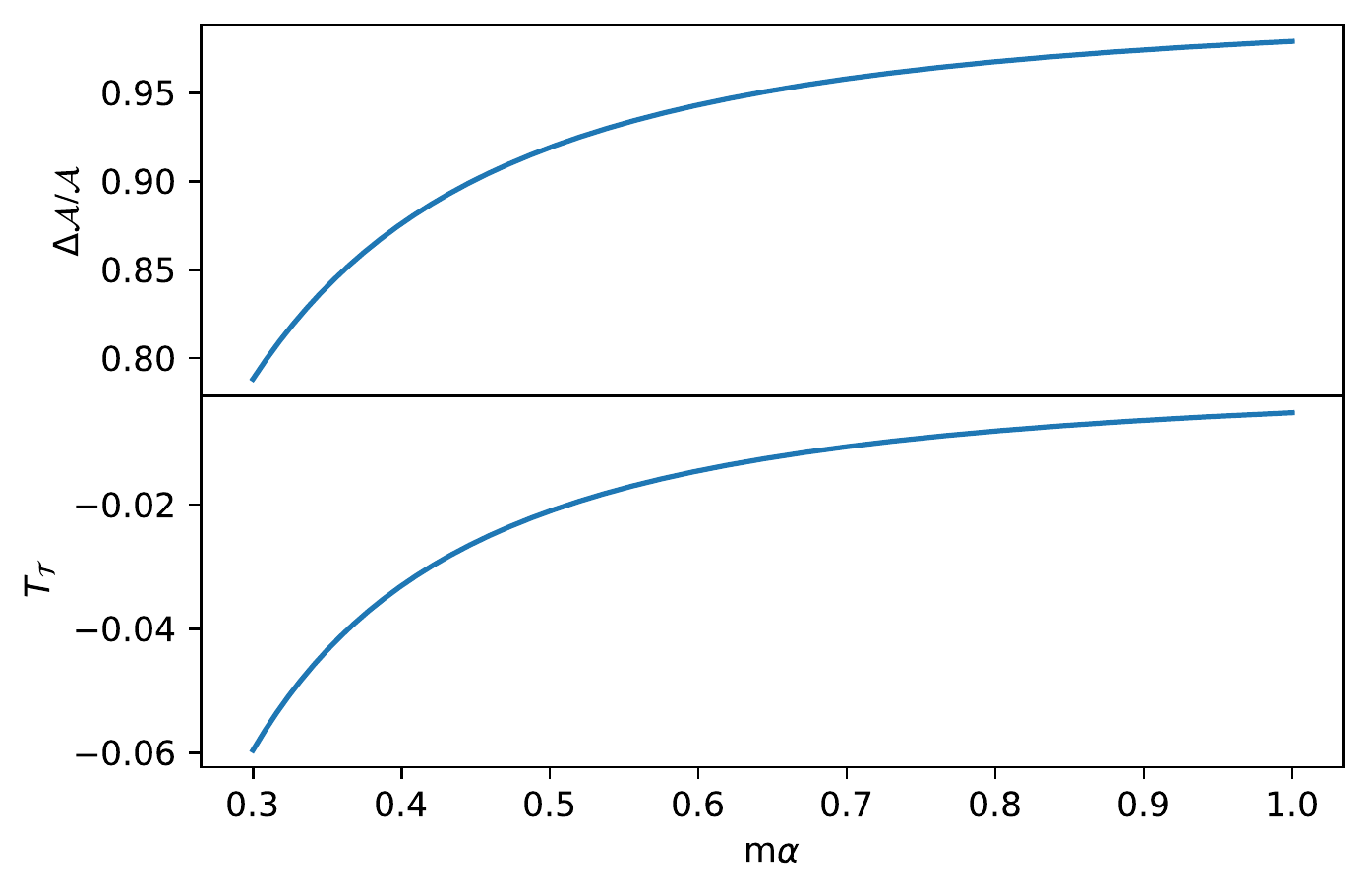}
 \end{center}
 \caption{Time of transition $T_{\cal T}$ and the relative correction to the horizon area as functions of $m$ for the intermediate roots. The panels correspond to the two branches with real solutions.}
 \label{plot3'}
 \end{figure}

\begin{figure}
 \begin{center}
\includegraphics[width=0.4\textwidth]{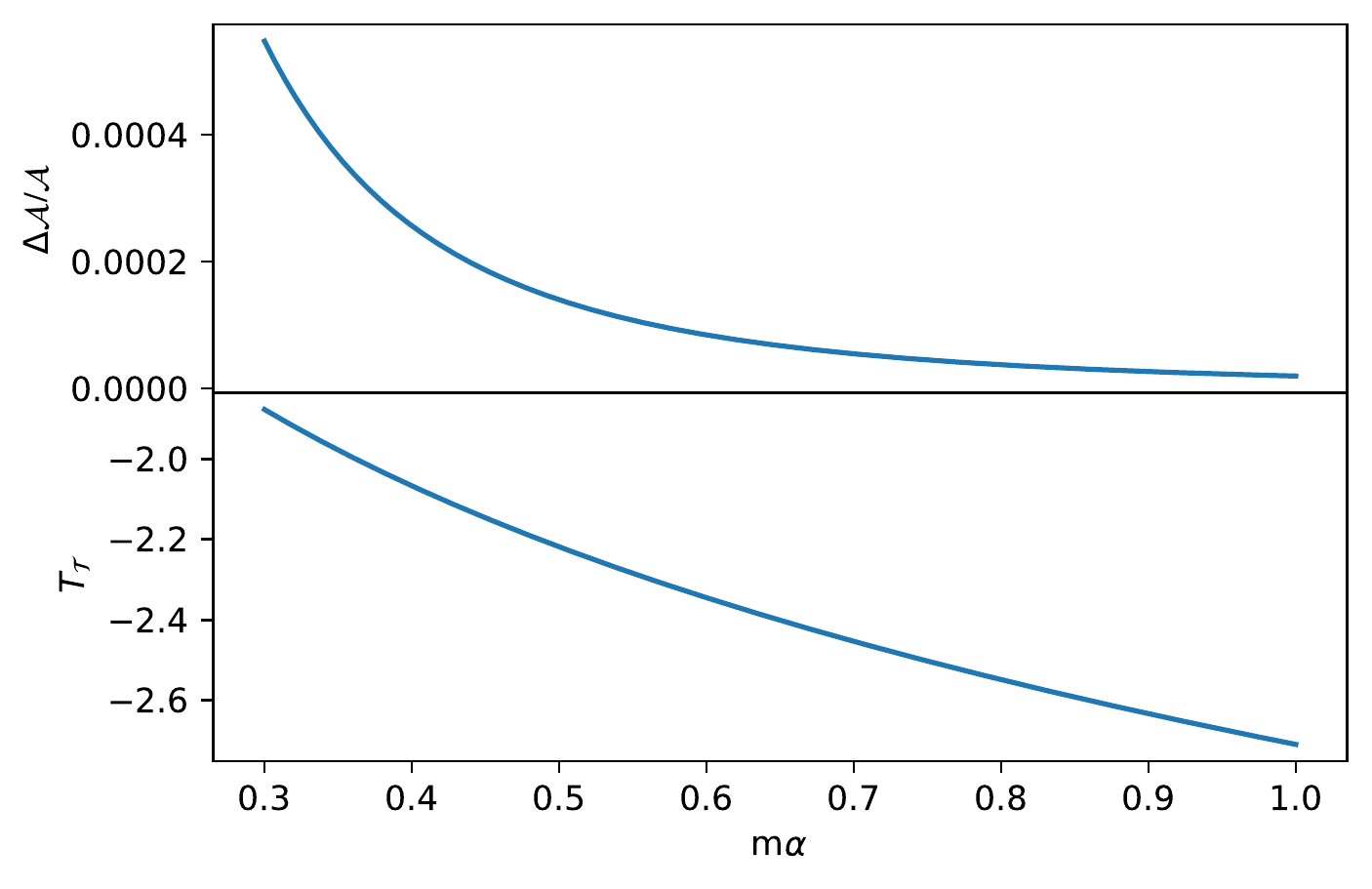}
 \end{center}
 \caption{Time of transition $T_{\cal T}$ and the relative correction to the horizon area as functions of $m$ for the physical root.}
 \label{plot5'}
 \end{figure}

 \begin{figure}
 \begin{center}
\includegraphics[width=0.4\textwidth]{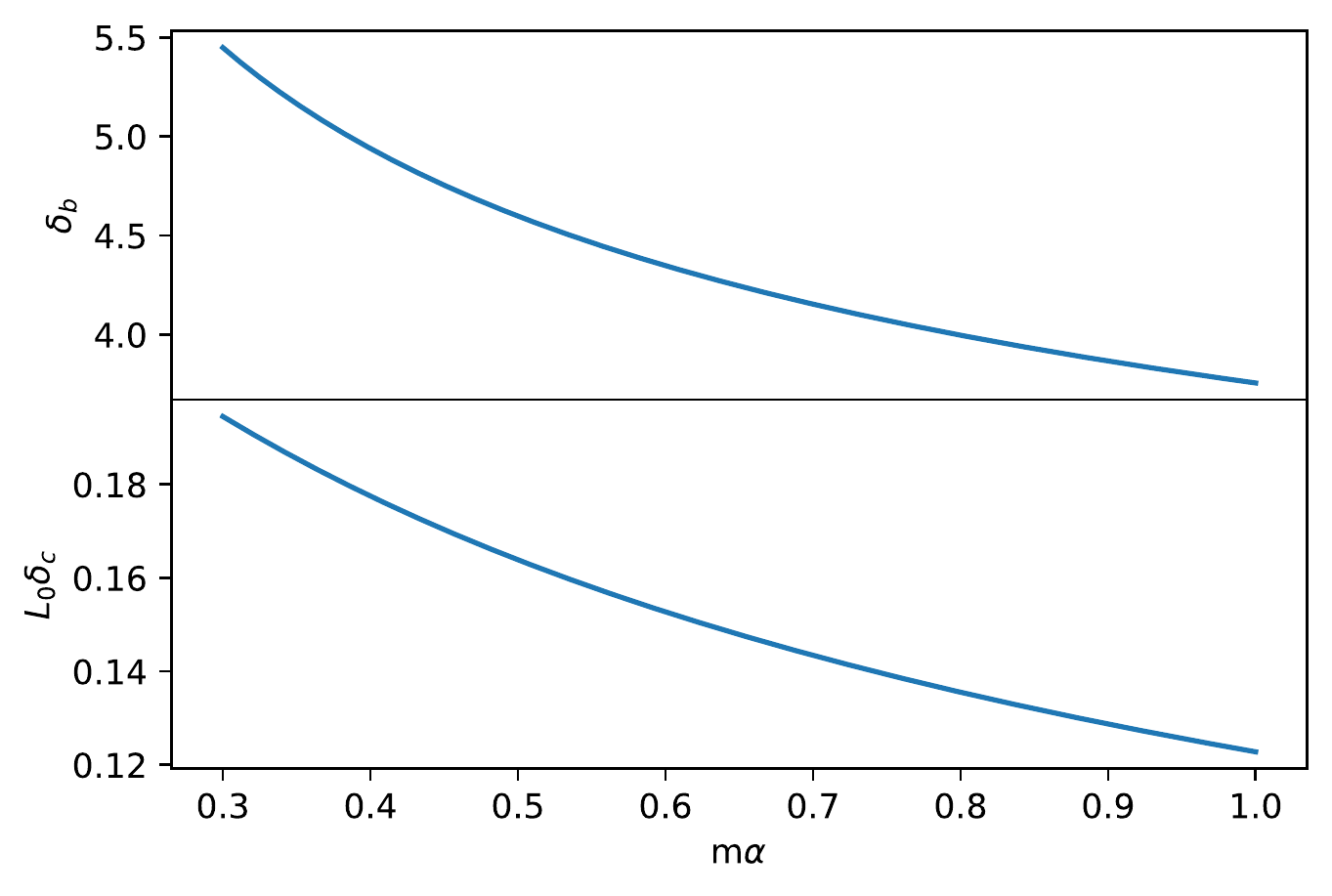}
 \end{center}
 \caption{Quantum parameters as functions of the black hole mass for the physical root.}
 \label{plot4'}
 \end{figure}

\subsection{Isolated horizons}

For Planck scale masses, evaporation effects could not, in principle, be neglected in the above computation. If we do not take into account for a moment the radiation backreaction on the metric, the quantum corrected Hawking temperature in the AOS model is given by \cite{Olmedo2}
\begin{equation}
T_H = \frac{1}{8\pi m} \frac{1}{(1+\epsilon_m)},
\end{equation}
where
\begin{equation}
\epsilon_m = \frac{\gamma^2 L_o^2 \delta_c^2}{64m^2}.
\end{equation}
With $\gamma = \sqrt{3}/6$, $m\alpha = 1$, $L_o \delta_c$ given by (\ref{1}) and the root for $\delta_b$ found above, we have $\epsilon_m \sim 10^{-3}$, and the Hawking temperature is still given by its large mass expression $T_H \approx 1/(8\pi m)$. Note that, due to (\ref{1}), the temperature correction $\epsilon_m$ is equal to the relative correction to the classical area in (\ref{horizon}). The smaller the latter, the smaller the former. Anyway, for a Planck mass black hole the temperature is about $4\%$ of the black hole mass, and the backreaction on the metric should also be estimated.

This difficulty can be circumvented if we consider an isolated horizon, as usually done in LQG. In this case, homogeneous slices of the Schwarzschild interior are not space-like as in Kantowski-Sachs spacetime, but time-like. This can be achieved through the replacements $b \rightarrow i \tilde{b}$ and $p_b \rightarrow i \tilde{p}_b$ \cite{ashtekar}, by which the lapse becomes
\begin{equation} \label{lapse2}
\tilde{N}_{\tau} = \frac{\gamma \sqrt{p_c} \delta_b}{\sinh(\delta_b \tilde{b})},
\end{equation}
and the interior metric assumes the form
\begin{equation} \label{KS2}
ds^2 = \tilde{N}_{\tau}^2 d\tau^2 - \frac{\tilde{p}_b^2}{p_c L_o^2} dx^2 + p_c (d\theta^2 + \sin^2\theta d\phi^2).
\end{equation}
Equations (\ref{m1}), (\ref{2.22}) and (\ref{Pc}) (and hence (\ref{4})) remain unaltered, while (\ref{2.25}) is rewritten as
\begin{equation} \label{2.25'}
\tilde{p}_b = - \frac{2\sin(\delta_c c)}{\delta_c} \frac{\sinh(\delta_b \tilde{b})}{\delta_b} \frac{p_c}{\gamma^2 - \frac{\sinh^2(\delta_b \tilde{b})}{\delta_b^2}}.
\end{equation}
Assuming that the constraints (\ref{4.5})-(\ref{4.4}) are still valid, we re-obtain (\ref{1}), Eq. (\ref{2}) changes sign\footnote{Now $\tilde{p}_b >0$ as in the corresponding classical solution \cite{ashtekar}.}, and (\ref{7}) is replaced by
\begin{equation} \label{7'}
\delta_b x^2 - x - \delta_b \gamma^2 = 0,
\end{equation}
where now
\begin{equation} \label{6'}
x \equiv \frac{\sinh (\delta_b \tilde{b})}{\delta_b}.
\end{equation}
Its roots lead to
\begin{equation} \label{8'}
\sinh(\delta_b \tilde{b}) = \frac{1 \pm \sqrt{1 + 4 \delta_b^2 \gamma^2}}{2} \quad \quad (\text{on}\; \cal{T}).
\end{equation}
Eq. (\ref{2.23}), by its turn, is replaced by
\begin{equation} \label{2.23'}
\cosh(\delta_b \tilde{b}) = b_o \tanh \left[ \frac{b_o T}{2} + \tanh^{-1} \left( \frac{1}{b_o} \right) \right],
\end{equation} 
with $T$ still given by (\ref{time}). For $m\alpha = 1$ and $\gamma = \sqrt{3}/6$, the largest root of the numerical solution of (\ref{8'})-(\ref{2.23'}) is $\delta_b \approx 3.75$. From (\ref{horizon}) we now have a relative correction to the horizon area of $\approx 2 \times 10^{-5}$. The numerical solution is presented in the lower panel of Fig.~\ref{plot1'}. There are two further non zero roots (shown in the zoom), very close to $\delta_b = 0.25$ ($0.249$ and $0.251$). For this value of $\delta_b$, the relative correction in (\ref{horizon}) is equal to $1$, and from (\ref{time}) and (\ref{1}) we have $T_{\cal T} = 0$, that is, the transition surface coincides with the horizon. For these roots (also present in the non-isolated case), quantum fluctuations are so large that the horizon is not formed. 

We performed the same analysis of isolated horizons varying the black hole mass in the interval $m\alpha \in [0.3,1]$, finding the roots shown in Fig.~\ref{plot2'}. The panels correspond to the two branches with real solutions. Apart $\delta_b = 0$, there are again two intermediate roots (upper and lower panels) with $m\alpha \delta_b \approx 0.25$, for which $T|_{\cal T} \approx 0$, and a largest root (upper panel). For the intermediate roots, we show in Fig.~\ref{plot3'} the dependence of the relative correction to the horizon area and of the time of transition $T_{\cal T}$ on the black hole mass. For one of the branches, the intermediate roots lead to $T_{\cal T} > 0$ and the horizon is not formed. For the other branch, the relative correction to the area increases with the mass, which is not an expected physical behaviour. Therefore, in this range of masses the only physical root is the largest one. In Fig.~\ref{plot5'} the relative correction to the horizon area for the physical roots is displayed as a function of mass, showing a decrease with $m$, while $T_{\cal T}$ remains negative, as expected. In Fig.~\ref{plot4'} we show that in this case both $\delta_b$ and $L_o \delta_c$ also decrease with the black hole mass. The qualitative behaviours shown in figures \ref{plot3'}-\ref{plot4'} are the same in the non-isolated horizon case (with different roots).

Note that, in our range of masses, $\delta_b > 1$ for the physical roots, which means that its interpretation as a physical fractional length on $\theta$-$\phi$ $2$-spheres (i.e. $\alpha = 1$) \cite{ashtekar} is not appropriate in this case. Actually, as $\delta_b$ decreases with the black hole mass, the value of $\alpha$ determines a minimal admissible mass, corresponding to $\alpha \delta_b = 1$ (when $4\pi p_c|_{\cal T}$ equals the area gap, see (\ref{4.5})). The relative correction to the horizon area for this minimal mass is given, from (\ref{horizon}), by
\begin{equation}
\frac{\Delta {\cal A}}{{\cal A}} = \frac{1}{256\, m^4}.
\end{equation}
If we assume, for example, that the minimal mass is the Planck mass (the mass of an extremal Kerr black hole with angular momentum $\hbar$), the relative correction to the horizon area is $\approx 4 \times 10^{-3}$. In this case the value of $\alpha$ corresponds to the intersection between the hyperbole $\delta_b = 1/(m\alpha)$ and the curve $\delta_b \times (m \alpha)$ shown in Fig.~\ref{plot4'}, and it is $\alpha \approx 0.13$, with $\delta_b \approx 7.5$. On the other hand, for the smallest Schwarzschild isolated horizon of LQG we have $m \approx 0.5,$\footnote{For two punctures with $j = 1/2$ we have $m \approx 0.5$, whereas a single puncture with $j = 1$ gives $m \approx 0.45$.} and the relative correction to the area is $\approx 0.06$.  As we can see, the Planck mass is a frontier between classical horizons, for which quantum corrections to the area are negligible, and quantum horizons for that they grow up to $6\%$.

For extremal Kerr black holes the Hawking radiation is totally null as the surface gravity is zero. The results obtained here for a spherically symmetric horizon do not apply to that case, of course. Nevertheless, the general conclusion is that quantum corrections to the area of Planck size horizons are not necessarily large, and may also be negligible for rotating black holes, as suggested by the precise correspondence shown above between Eqs. (\ref{LQG1'}) and (\ref{LQG1}). Although the AOS model is hardly generalisable to rotating black holes, its extension to extremal charged horizons is straightforward, as we show below. We will see that relative corrections to the horizon area are small (though not negligible) also in this case.

\subsection{Extremal charged horizons}

The extremal Reisnner-Nordstr\"om metric (for which $Q = m$) can be written as
\begin{equation}
ds^2 = -\left( 1 - \frac{m}{r} \right)^2 dt^2 + \left( 1 - \frac{m}{r} \right)^{-2} dr^2 + r^2 d\Omega,
\end{equation}
where $d\Omega = d\theta^2 + \sin^2\theta\, d\phi^2$. The black hole interior can be described through the variable changes $r \rightarrow \tau$, $t \rightarrow x$ \cite{ashtekar}, under which the metric is rewritten as
\begin{equation}
ds^2 = \left( \frac{m}{\tau} - 1 \right)^{-2} d\tau^2 - \left( \frac{m}{\tau} - 1 \right)^2 dx^2 +  \tau^2 d\Omega.
\end{equation}
If we now use the dictionary
\begin{eqnarray}
\tilde{N}_{\tau}^2 &=& \left( \frac{m}{\tau} - 1 \right)^{-2},\\
\tilde{p}_b^2 &=& L_o^2 \left( \frac{m}{\tau} - 1 \right)^2 \tau^2,\\
p_c &=& \tau^2,
\end{eqnarray}
the metric assumes the form (\ref{KS2}) of an isolated horizon, with the difference that, now, the horizon corresponds to $\tau = m$. This result assures that the quantum corrected horizon does not radiate either, which is essential for the identification of an eigenstate of the area operator. Whether this is also valid for an extremal rotating black hole is something deserving further investigation. For the extremal charged horizon, all the results of last section apply, with the change $m \rightarrow m/2$. For the horizon area we have
\begin{equation} \label{horizon2}
{\cal A} = 4 \pi m^2 \left( 1 + \frac{3 \gamma^2}{4 m^4 \alpha^4 \delta_b^4} \right).
\end{equation}
The minimal admissible mass corresponds again to $\delta_b \alpha = 1$. For $\gamma = \sqrt{3}/6$, an isolated horizon pierced twice by a line with $j = 1/2$ has $m \approx 1$ in this case\footnote{For a single puncture with $j = 1$ we have $m\approx 0.9$.}. Therefore, from (\ref{horizon2}) we have $\delta{\cal A}/{\cal A} \approx 0.06$. We note that for both the Schwarzschild isolated horizon and the charged extremal horizon, for the minimal masses the relative correction to the area is $\approx 6\%$, whereas for the extremal rotating black hole it is an order of magnitude smaller. This is in fact expected, as in the former cases the minimal area corresponds to two area gaps, while in the later it corresponds to four punctures (see (\ref{LQG1})). For a given irreducible representation, the larger the number of lines piercing the horizon, the smaller the relative correction to the classical area. 

The reader may ask whether a model with spherical symmetry can be used at so small scales. Actually, effective models with homogeneous metrics have also been used at the Planck scale in Loop Quantum Cosmology for resolving the big-bang singularity \cite{LQC2,LQC}. Furthermore, even when dealing with large black holes, the AOS model makes use of the Kantowski-Sachs metric at the transition surface where quantum effects are not negligible. The use of these approximate models at such scales should be rather verified by evaluating the corrections they predict for classical quantities like the horizon area. Deep inside the horizon quantum effects are surely large, so large that the central singularity disappears. Nevertheless, the horizon remains almost unaffected, what permits the use of the classical theory in the evaluation of its mass and spin, as above.

\subsection{Macroscopic horizons}

\begin{figure}
 \begin{center}
\includegraphics[width=0.4\textwidth]{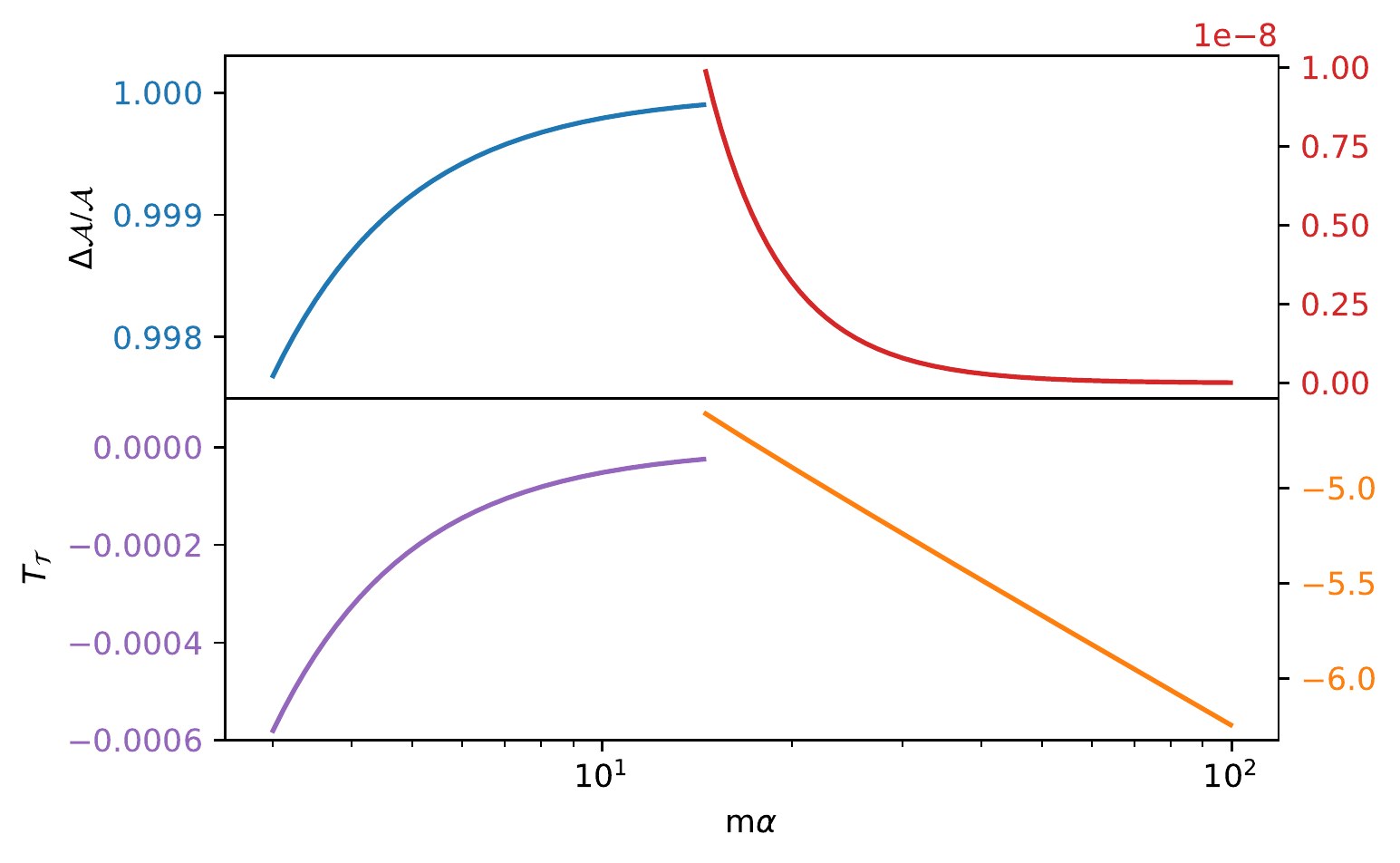}
 \end{center}
 \caption{Time of transition $T_{\cal T}$ and the relative correction to the horizon area as functions of $m$ for one of the intermediate roots in the non-isolated case.}
 \label{macroscopic}
 \end{figure}

An interesting difference of the above analysis with respect to the original AOS model is that the latter treats with macroscopic black holes, although making use of large quantum corrections at the transition surface to avoid the singularity. In Ref. \cite{ashtekar}, the authors find four roots for $\delta_b$ as above, namely zero, two intermediate (and approximately equal) roots and a larger one, with slightly different numerical values due to the value adopted for $\gamma$ and the non-isolated character of the horizons. However, they dismiss the larger root since it leads to unphysical results in this range of masses, e.g. large quantum corrections in low curvature regions. On the other hand, the intermediate roots present a physical behaviour, contrary to what we have found here. 

In order to understand this apparent contradiction, we plot in Fig.~\ref{macroscopic} the evolution of $\Delta{\cal A}/{\cal A}$ and $T_{\cal T}$ with the black hole mass for one of the intermediate roots of the non-isolated horizon case, when the mass ranges up to macroscopic values (for the other roots the scalings are the same as shown in Fig.~\ref{plot5'} and in the upper panel of Fig.~\ref{plot3'}). For masses below the Planck scale, we reproduced the unphysical scalings shown in the lower panel of Fig.~\ref{plot3'}, with quantum corrections increasing with the mass. For large masses we have, however, a physical behaviour, with the corrections decreasing fast with $m$. The figure also evidences a discontinuity at the Planck scale.  We see, therefore, that macroscopic and microscopic black holes seem to belong to disconnected domains, with the former being described by one of the intermediate roots while the latter are described by the larger one. That discontinuity suggests that, when the mass of a macroscopic black hole decreases (owing, for example, to an adiabatic evaporation process), the horizon is disrupted by large quantum fluctuations when its mass approaches the Planck scale. If this is true, microscopic horizons can only be formed in collision processes like that hypothetically considered in Section~\ref{Section II}. The real conditions for such collisions to form actual horizons are not well understood (in particular for the case of our two neutrinos black hole). The reader can find a detailed discussion on this subject in Ref. \cite{carr}.\footnote{See also \cite{'tHooft}. A study on collisions of relativistic black holes can be found in \cite{new}.}

\section{Concluding remarks}

We have obtained three curious numerical coincidences in this paper. Using classical GR, we have shown that a black hole formed by parallel neutrinos in the lowest-mass massive state $m_2$ (assuming $m_1 = 0$) is an extremal Kerr black hole with Planck mass, within $99.9\%$ precision. Its classical horizon is eigenstate of the LQG area operator, provided that the Barbero-Immirzi parameter is $5\%$ above the large area approximate value. And, with the new BI parameter, we have precisely obtained (less than $1\%$ error) the Bekenstein-Hawking slope for the leading term of the entropy vs area relation for Planck size horizons.
In the absence of a full LQG solution for the Kerr black hole or a compelling effective quantum model that includes matter (but see e.g. \cite{modesto,modesto2}),
in the derivations above we assumed that Planck size black holes have classical horizons, that is, quantum corrections are negligible at the Planck scale. This assumption seems corroborated by our entropy analysis and was verified within an effective model for spherically symmetric quantum black holes.

The eigenstate suggested here has implications that transcend a possible signature of quantum gravity and the determination of the BI parameter. Some assumptions were made and finding $m_2$ with that precision may be a consistency test for those assumptions. In the gravity sector we have assumed, as already pointed out, that quantum gravity corrections to the black holes horizon area and to neutrinos magnetic moments are negligible above the Planck length. We have also taken for granted that the scattered magnetic dipoles are no longer observed from outside after the horizon formation, leading to a Kerr solution characterised only by $J$ and $M$, as postulated by the classical no-hair conjecture. Furthermore, the value found for the BI parameter agrees with the Ghosh-Mitra count of states \cite{mitra}, in opposition to the Dogamala-Lewandowski original count \cite{DL}. These two counts lead to the same ${\cal S} \times {\cal A}$ relation in the large area limit, but with different values for $\tilde{\gamma}$ in Eq. (\ref{Bekenstein}) \cite{meissner}.   Nevertheless, the most intriguing implication is perhaps the corroboration of some assumptions made in the neutrinos sector. Expression (\ref{dipoles4}) for the neutrinos magnetic moment is only valid if they are Dirac neutrinos, because Majorana neutrinos do not carry magnetic moments. In its derivation it is also assumed the minimal extension of the Standard Model needed to accommodate massive neutrinos, with the addition of right-handed singlets  \cite{dipole}. Finally, the assumption of normal ordering of the mass states, with $m_1 = 0$, was necessary. We see that looking for empirical validation of quantum gravity may shed light on other, apparently uncorrelated open questions.

Nevertheless, the results presented here may be considered provisory in view of some difficult issues still open. First of all, we must verify that the extremal rotating black hole is solution of the Hamiltonian constraint of LQG. If it does not radiate it represents a stationary state, but this is hard to verify even in the realm of effective models, designed to describe spherically symmetric holes. Its identification as an area operator eigenstate rests exclusively on the correspondence between the classical horizon area and the area operator eigenvalues, correspondence explored with the help of the AOS formalism. Although inspired in LQG, our results may rather be considered effective approximations, still requiring verification on a more fundamental ground.

On the other hand, the small discrepancy between our value of the BI parameter and the Gosh-Mitra value evidences the approximate derivation of the latter, that uses the entropy of large horizons. The numerical results of Section III, that represents the kern of this paper, are robust and do not depend on assumptions made in other sections. We have followed the recipe given in Refs.~\cite{agullo,corichi}, among others, to obtain the entropy $\times$ area relation from the LQG area operator eigenvalues. The results of Figure 2 suggest that the approximate value of the BI parameter found by Meissner \cite{meissner} on the basis of the Gosh-Mitra count \cite{mitra} using large horizons is indeed $5\%$ below the actual value.

\section*{Acknowledgements}

We are thankful to G. A. Mena Marug\'an and J. Olmedo for valuable comments and suggestions, and to J. C. Fabris, P. C. de Holanda, O. L. G. Peres, A. Saa and J. Zanelli for useful discussions. S.C. is partially supported by CNPq (Brazil) with grant No.  307467/2017-1.

\end{document}